\global\def\draftcontrol{0}

   \def\versionno{ cascading quasinormal  modes}

\catcode`\@=11

\expandafter\ifx\csname draftcontrol\endcsname\relax\global\def\draftcontrol{0}
\fi

{\count255=\time\divide\count255 by 60
\xdef\hourmin{\number\count255}
\multiply\count255 by-60\advance\count255 by\time
\xdef\hourmin{\hourmin:\ifnum\count255<10 0\fi\the\count255}}
\def\draftdate{\number\month/\number\day/\number\year\ \ \ \hourmin }

\newcommand\makepapertitle{\par
  \begingroup
    \renewcommand\thefootnote{\@fnsymbol\c@footnote}%
    \def\@makefnmark{\rlap{\@textsuperscript{\normalfont\@thefnmark}}}%
    \long\def\@makefntext##1{\parindent 1em\noindent
            \hb@xt@1.8em{%
                \hss\@textsuperscript{\normalfont\@thefnmark}}##1}%
     \newpage
     \global\@topnum\z@   
     \@makepapertitle
     \thispagestyle{empty}\@thanks
  \endgroup
  \setcounter{footnote}{0}%
  \global\let\thanks\relax
  \global\let\makepapertitle\relax
  \global\let\@makepapertitle\relax
  \global\let\@thanks\@empty
  \global\let\@author\@empty
  \global\let\@date\@empty
  \global\let\@title\@empty
  \global\let\title\relax
  \global\let\author\relax
  \global\let\date\relax
  \global\let\and\relax
  \def\version{\let\version\@version\@gobble}
}
\def\@makepapertitle{%
  \newpage
   \ifnum\draftcontrol=1 {}
   \version\versionno
   \vskip 3em%
   \else
   \hfill\hbox to 3cm {\parbox{4cm}{\@pubnum}\hss}%
   \vskip 3em%
   \fi
   \begin{center}%
   \let \footnote \thanks
     {\LARGE {\@title}}%
     \vskip 1.5em%
     {\normalsize
       \lineskip .5em%
       \begin{tabular}[t]{c}%
         \@author
       \end{tabular}\par}%
     \vskip 1.5em%
     {\@bstract}%
     \end{center}%
     \vskip 1.5em
     \@date%
   \par
}

\gdef\@pubnum{}
\def\pubnum#1{%
  \gdef\@pubnum{#1}}

\gdef\@bstract{}
\def\Abstract#1{%
  \gdef\@bstract{%
   \parbox{\textwidth-0pc}{%
   \centerline{\bf Abstract}\penalty1000%
\kern.2cm%
\noindent
\renewcommand\baselinestretch{1.0}%
{#1}}}
}

\def\ps@paper{\let\@mkboth\@gobbletwo%
     \ifnum\draftcontrol=1
    \def\@oddfoot{\hbox to \textwidth{\tiny \versionno \hfil\tiny\draftdate}%
    \hskip -\textwidth \hbox to \textwidth{\hfil\rm\thepage\hfil}}%
     \else\def\@oddfoot{\hbox to \textwidth{\hfil\rm\thepage\hfil}}
     \fi
     \let\@evenfoot\@oddfoot
}

\def\body{\clearpage
          \pagestyle{paper}
    }

\def\@version#1{\ifnum\draftcontrol=1
\typeout{}\typeout{#1}\typeout{}
\vskip3mm\centerline{\hbox{\fbox{\normalsize{\tt DRAFT -- #1 -- }
                   {\draftdate}}}}\vskip3mm
\fi}
\let\version\@version
\long\def\eqlabel#1{\ifnum\draftcontrol=1
                    \tag@false  
                    \tag*{(\theequation) \hbox to -0.2cm{\hspace{0cm}\small{#1}\hss}}
                    \refstepcounter{equation}
                    \edef\@currentlabel{\theequation}
                    \ltx@label{#1}          
                    \else
                    \label{#1}
                    \fi
                    }
\let\st@bibitem\@bibitem
\let\st@lbibitem\@lbibitem
\ifnum\draftcontrol=1
  \def\@bibitem#1{%
    \st@bibitem{#1}\a@@label{#1}\ignorespaces}
  \def\@lbibitem[#1]#2{%
    \st@lbibitem[#1]{#2}\a@@label{#2}\ignorespaces}
  \def\a@@label#1{%
    \gdef\a@lab{\smash{\normalfont\small#1}}
    \ifvmode
      \if@inlabel
        \global\setbox\@labels\hbox{%
          \llap{\a@lab\let\a@lab\relax
                \kern\@totalleftmargin\kern\marginparsep}%
          \box\@labels}%
      \fi
    \fi}
\fi

\documentclass[12pt,letterpaper]{article}

\usepackage{amsmath,amssymb,array,calc,epsfig,rotating,bm}
\usepackage[sort]{cite}
\usepackage{graphicx}
\usepackage{psfrag,verbatim}

\ifnum\draftcontrol=1
\fi

\renewcommand\baselinestretch{1.25}
\renewcommand\section{\@startsection {section}{1}{\z@}%
                                   {-3.5ex \@plus -1ex \@minus -.2ex}%
                                   {2.3ex \@plus.2ex}%
                                   {\normalfont\large\bfseries}}
\renewcommand\subsection{\@startsection{subsection}{2}{\z@}%
                                   {-3.25ex\@plus -1ex \@minus -.2ex}%
                                   {1.5ex \@plus .2ex}%
                                   {\normalfont\normalsize\bfseries}}
\renewcommand\subsubsection{\@startsection{subsubsection}{3}{\z@}%
                                   {-3.25ex\@plus -1ex \@minus -.2ex}%
                                   {1.5ex \@plus .2ex}%
                                   {\normalfont\normalsize\it}}
\renewcommand\paragraph{\@startsection{paragraph}{4}{\z@}%
                                   {-3.25ex\@plus -1ex \@minus -.2ex}%
                                   {1.5ex \@plus .2ex}%
                                   {\normalfont\normalsize\bf}}

\numberwithin{equation}{section}

\def\revise#1       {\raisebox{-0em}{\rule{3pt}{1em}}%
                     \marginpar{\raisebox{.5em}{\vrule width3pt\
                     \vrule width0pt height 0pt depth0.5em
                     \hbox to 0cm{\hspace{0cm}{%
                     \parbox[t]{4em}{\raggedright\footnotesize{#1}}}\hss}}}}

\newcommand{\ie}{{\it i.e.,}\ }

\def\caln         {{\cal N}}
\def\calo         {{\cal O}}

\def\zet          {{\mathbb Z}}

\def\tr           {\mathop{\rm Tr}}
\def\Re           {{\rm Re\hskip0.1em}}
\def\Im           {{\rm Im\hskip0.1em}}

\def\sqr#1#2{{\vcenter{\vbox{\hrule height.#2pt
 \hbox{\vrule width.#2pt height#1pt \kern#1pt
 \vrule width.#2pt}\hrule height.#2pt}}}}

\newcommand{\ft}[2]{{\textstyle{\frac{#1}{#2}}}}

\def\w{\omega}

\def\dd{\delta}
\def\e{\epsilon}
\def\c{\chi}

\def\g{\gamma}

\def\aa1{\phi}
\def\cc1{\psi}

\catcode`\@=12

\begin{document}


\title{\bf   Universal relaxation in quark-gluon plasma at strong coupling}

\date{May 19, 2015}

\author{
Alex Buchel$^{1,2,3}$ and Andrew Day$^{1}$ \\[0.4cm]
\it $^1$\,Department of Applied Mathematics, $^2$\,Department of Physics and Astronomy, \\
\it University of Western Ontario\\
\it London, Ontario N6A 5B7, Canada\\
\it $^3$\,Perimeter Institute for Theoretical Physics\\
\it Waterloo, Ontario N2J 2W9, Canada
}

\Abstract{We use top-down gauge theory/string theory correspondence to compute
relaxation rates in strongly coupled nonconformal gauge theory
plasma. We compare models with difference mechanisms of breaking the
scale invariance: "hard breaking'' (by relevant operators) and "soft
breaking'' (by marginal operators).  We find that the thermalization
time of the transverse traceless fluctuations of the stress-energy
tensor is rather insensitive to the mechanisms of breaking the scale
invariance over a large range of the scale-breaking parameter
$\delta=\frac 13-c_s^2$. We comment on the relevance of the results to
QCD quark-gluon plasma.
}

\makepapertitle

\body

\version\versionno
\tableofcontents

\section{Introduction}\label{intro}

In a recent paper \cite{Buchel:2015saa} it was pointed out that the equilibration rates 
in strongly coupled nonconformal quark-gluon plasma (QGP) are surprisingly 
insensitive to the presence of the conformal symmetry breaking scale. 
Specifically, considering  supersymmetric mass deformations  within 
$\caln=2^*$ gauge theory and using holography \cite{m1,Aharony:1999ti} the authors computed the 
spectra of quasinormal modes for a variety of scalar operators, as well as the energy-momentum tensor. In each case, 
the lowest quasinormal frequency, which provides an approximate upper bound on the thermalization time, 
was found to be proportional to temperature, up to a pre-factor with only a mild temperature dependence. 
Similar results were reported for a conformal plasma with a finite charge density and in the 
presence of an  external magnetic field \cite{Fuini:2015hba}, as well
as in a phenomenological nonconformal holographic models  \cite{Janik:2015waa,Ishii:2015gia}.

In this paper we continue investigation of the equilibration rates in 
strongly coupled quark-gluon plasma with holographic string 
theory dual (the top-down models). One drawback of $\caln=2^*$ model 
studied in \cite{Buchel:2015saa} is the fact that the conformal invariance 
there is broken quite mildly --- over the range of the supersymmetric 
mass deformation parameter $\frac{m}{T}$, the scale invariance is 
violated by  
\begin{equation}
\max_{\ft mT} \frac{\e-3p}{\e} \approx 20\%\,.
\eqlabel{n2trace}
\end{equation}
In \eqref{n2trace} $\e$ and $p$ are the energy density and the pressure in $\caln=2^*$ plasma. 
On the contrary, the latest results from the HotQCD Collaboration \cite{Bazavov:2014pvz}
indicate that the analogous quantity in QCD is approximately $50\%$. Furthermore, 
it is important to verify how robust are the results of \cite{Buchel:2015saa} 
in theories with a different mechanism for breaking the scale invariance. 
The ideal model to address these two questions is the Klebanov-Strassler (KS) 
cascading gauge theory \cite{ks}. First, the conformal invariance in 
KS gauge theory is broken much stronger \cite{kstalk}; second, while the renormalization group (RG)
flow in $\caln=2^*$ gauge theory is induced by relevant operators, the RG flow in KS gauge theory is induced 
by  marginal, but not exactly marginal, operators. We compute the lowest quasinormal mode 
associated with the transverse traceless fluctuations of the stress-energy tensor in KS gauge theory.

We omit technical details and focus on results only\footnote{The thermodynamics of 
KS gauge theory has been studied in  \cite{b,kbh1,kbh2,aby,abk,hyd3,ksbh}.}. 
In the next section we recall definitions of $\caln=2^*$ gauge theory and
KS gauge theory. We compare the thermodynamics of the two models with 
that of the lattice QCD \cite{Bazavov:2014pvz}. In section \ref{quasi}
we present results for the lowest quasinormal mode of the transverse 
traceless fluctuations of the stress-energy tensor in KS plasma, and compare them 
with the corresponding computations in \cite{Buchel:2015saa}. 
Finally, we conclude in section \ref{conclude}.

\section{Thermodynamics of strongly coupled nonconformal plasma from holography}\label{thermo}

The best studied example of the gauge theory/string theory correspondence is that between 
the maximally supersymmetric $\caln=4$ $SU(N)$ supersymmetric Yang-Mills theory (SYM)
and string theory in $AdS_5\times S^5$ \cite{m1}. SYM is conformally invariant. 
At strong coupling, the energy density and the pressure of equilibrium SYM plasma 
at temperature $T$ is given by 
\begin{equation}
\e=\frac{3}{8}\pi^2 N^2 T^4\,,\qquad p=\frac{1}{8}\pi^2 N^2 T^4\,.
\eqlabel{n4}
\end{equation}  

In what follows we find it convenient to parameterized thermodynamic potentials of nonconformal plasma with the following conformal symmetry breaking 
parameters:
\begin{equation}
\Theta\equiv \frac{\e-3p}{\e}\,,\qquad \dd\equiv \frac 13-c_s^2\,,
\eqlabel{cftbreaking}
\end{equation}
where $c_s$ is the speed of sound waves in plasma. As we see below, parameterization \eqref{cftbreaking} allows to compare different 
holographic models with lattice QCD.  Note that from \eqref{n4}, 
\begin{equation}
\Theta\bigg|_{\caln=4}=0\,,\qquad \dd\bigg|_{\caln=4}=0\,.
\eqlabel{n4trace}
\end{equation}

$\caln=2^*$ gauge theory is obtained as a mass deformation of $\caln=4$ SYM, where an $\caln=2$ hypermultiplet receives 
a mass $m$. This is a {\it relevant} deformation of the conformal SYM, as the renormalization group flow is induced 
by bosonic and fermionic mass terms of the hypermultiplet. At large temperatures, \ie $\frac{m}{T}\ll 1$,
the thermodynamics of $\caln=2^*$ gauge theory plasma is given by \cite{Buchel:2003ah,Buchel:2004hw}
\begin{equation}
\begin{split}
&\e=\frac{3}{8}\pi^2 N^2 T^4\biggl( 1-\frac 23\frac{\Gamma(3/4)^4}{\pi^4}\frac{m^2}{T^2}+\calo\left(\frac{m^4}{T^4}\ln\frac Tm\right)\biggr)\,,\\
&p=\frac{1}{8}\pi^2 N^2 T^4\biggl( 1-2\frac{\Gamma(3/4)^4}{\pi^4}\frac{m^2}{T^2}+\calo\left(\frac{m^4}{T^4}\ln\frac Tm\right)\biggr)\,,
\end{split}
\eqlabel{n2thermo}
\end{equation}
resulting in 
\begin{equation}
\Theta=6\dd +\calo(\dd^2\ln\dd)\,.
\eqlabel{thetan2}
\end{equation}

Klebanov-Strassler cascading gauge theory is $\caln=1$ four-dimensional supersymmetric $SU(K+P)\times SU(K)$
gauge theory with two chiral superfields $A_1, A_2$ in the $(K+P,\overline{K})$
representation, and two fields $B_1, B_2$ in the $(\overline{K+P},K)$ representation.
Perturbatively, this gauge theory has two gauge couplings $g_1, g_2$ associated with 
two gauge group factors,  and a quartic 
superpotential
\begin{equation}
W\sim \tr \left(A_i B_j A_kB_\ell\right)\e^{ik}\e^{j\ell}\,.
\eqlabel{w}
\end{equation}
When $P=0$ above theory flows in the infrared to a 
superconformal fixed point, commonly referred to as Klebanov-Witten (KW) 
theory \cite{kw}. At the IR fixed point KW gauge theory is 
strongly coupled --- the superconformal symmetry together with 
$SU(2)\times SU(2)\times U(1)$ global symmetry of the theory implies 
that anomalous dimensions of chiral superfields $\gamma(A_i)=\gamma(B_i)=-\frac 14$, \ie non-perturbatively large.
Notice that the superpotential \eqref{w} is {\it marginal} at the fixed point. 
When $P\ne 0$, conformal invariance of the above $SU(K+P)\times SU(K)$
gauge theory is broken. It is useful to consider an effective description 
of this theory at energy scale $\mu$ with perturbative couplings
$g_i(\mu)\ll 1$. It is straightforward to evaluate NSVZ beta-functions for 
the gauge couplings. One finds that while the sum of the gauge couplings 
does not run
\begin{equation}
\frac{d}{d\ln\mu}\left(\frac{\pi}{g_s}\equiv \frac{4\pi}{g_1^2(\mu)}+\frac{4\pi}{g_2^2(\mu)}\right)=0\,,
\eqlabel{sum}
\end{equation}
the difference between the two couplings is  
\begin{equation}
\frac{4\pi}{g_2^2(\mu)}-\frac{4\pi}{g_1^2(\mu)}\sim P \ \left[3+2(1-\g_{ij})\right]\ \ln\frac{\mu}{\Lambda}\,,
\eqlabel{diff}
\end{equation}
where $\Lambda$  is the strong coupling scale of the theory and $\g_{ij}$ are anomalous dimensions of operators $\tr A_i B_j$.
Given \eqref{diff} and \eqref{sum} it is clear that the effective weakly coupled description of $SU(K+P)\times SU(K)$ gauge theory 
can be valid only in a finite-width energy band centered about $\mu$ scale. Indeed, extending effective description both to the UV 
and to the  IR one necessarily encounters strong coupling in one or the other gauge group factor. As  explained 
in \cite{ks}, to extend the theory past the strongly coupled region(s) one must perform  Seiberg duality \cite{sd}. 
Turns out, in this gauge theory,  Seiberg duality transformation is a self-similarity transformation of the effective description 
so that $K\to K-P$ as one flows to the IR, or $K\to K+P$ as the energy increases. Thus, extension of the effective 
$SU(K+P)\times SU(K)$ description to all energy scales involves and infinite sequence - a {\it cascade } - of Seiberg dualities
where the rank of the gauge group is not constant along RG flow, but changes with energy according to \cite{b} 
\begin{equation}
K=K(\mu)\sim 2 P^2 \ln \frac \mu\Lambda\,, 
\eqlabel{effk}
\end{equation}
at least as $\mu\gg \Lambda$. Since \cite{ks}  
\begin{equation}
\g_{ij}=-\frac 12+\calo\left(\frac{P^2}{K^2}\right)\,,  
\eqlabel{guv}
\end{equation}
the superpotential \eqref{w} is marginal, but {\it not exactly marginal} at the 
Klebanov-Witten ultraviolet fixed point of the theory. 
Although there are infinitely many duality cascade steps in the UV, there is only a finite number of duality transformations as one 
flows to the IR (from a given scale $\mu$). The space of vacua of a generic cascading gauge 
theory was studied in details in 
\cite{dks}. In the simplest case, when $K(\mu)$ is an integer multiple of $P$, cascading gauge 
theory confines in the 
infrared with a spontaneous breaking of the chiral symmetry $U(1)\supset \zet_2$ \cite{ks}. 
Here, the full global symmetry of the ground state is $SU(2)\times SU(2)\times \zet_2$.  
At large temperatures, \ie $\frac{\Lambda}{T}\ll 1$,  the thermodynamics of 
KS gauge theory plasma is given by 
\begin{equation}
\begin{split}
&\e=\frac{243}{256}\frac{\Lambda^4}{\pi^4}e^{\frac{2 K(T)}{P^2}}\left(1+\frac{P^2}{3 K(T)}+\calo\left(\frac{P^4}{K(T)^2}\right)\right)
\,,\\
&p=\frac{81}{256}\frac{\Lambda^4}{\pi^4}e^{\frac{2 K(T)}{P^2}}\left(1-\frac{P^2}{ K(T)}+\calo\left(\frac{P^4}{K(T)^2}\right)\right)\,,
\end{split}
\eqlabel{ksthermo}
\end{equation}
with 
\begin{equation}
\frac{dK(T)}{d\ln \frac{T}{\Lambda}}=2P^2 \left(1+\calo\left(\frac{P^2}{K(T)}\right)\right)\,, 
\eqlabel{dkdt}
\end{equation}
resulting in 
\begin{equation}
\Theta=3\dd +\calo(\dd^2)\,.
\eqlabel{thetaks}
\end{equation}

\begin{figure}[t]
\begin{center}
\psfrag{x}{{$\dd$}}
\psfrag{y}{{$\Theta$}}
\includegraphics[width=5in]{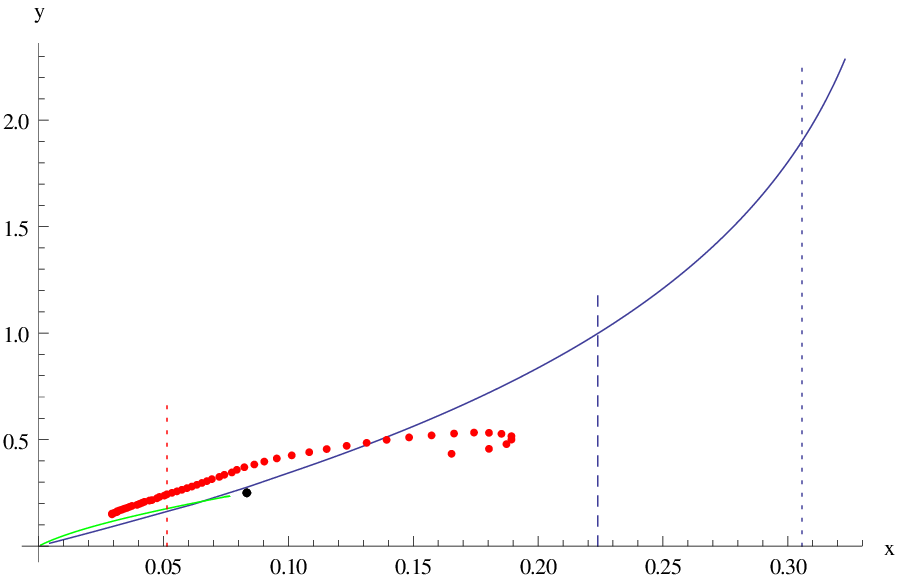}
\end{center}
  \caption{ Parameterization of $\Theta=\frac{\e-3p}{\e}$ with $\dd=\frac 13-c_s^2$ in strongly coupled 
gauge theory plasma for QCD (the red dots),  $\caln=2^*$ (the solid green line), 
and  cascading gauge theory (the solid blue line).  The dashed red line represents the conformal violation parameter 
$\dd$ in QCD at $T=0.3$GeV.  The black dot is the $\frac{m}{T}\to 0$ limit of  $\caln=2^*$
thermodynamics. Vertical blue lines represent the phase transitions in cascading gauge theory plasma:
the confinement/deconfinement (dashed) and the chiral symmetry breaking (dotted).} \label{figure1}
\end{figure}

Using results of Table I of \cite{Bazavov:2014pvz} we can reconstruct $\Theta$-vs-$\dd$ for QCD. 
These results are presented by red dots in figure \ref{figure1}. The dashed vertical red line 
represents QCD nonconformality parameter $\dd$ at temperature of\footnote{We choose this as a characteristic temperature for 
initializing hydrodynamic codes to model RHIC collisions.} 0.3GeV.  QCD data points at temperatures higher than 
0.3GeV correspond to weaker breaking of conformal invariance --- they are to the left of the dashed red line. 
The solid green line parameterizes $\caln=2^*$ thermodynamics \cite{Buchel:2007vy}. 
In the deep infrared, \ie $\frac{m}{T}\to 0$, 
$\caln=2^*$ thermodynamics reduces to that of the five-dimensional CFT \cite{Buchel:2007mf,HoyosBadajoz:2010td}. The latter 
limit is represented by a black dot,
\begin{equation}
\{\dd,\Theta\}\bigg|_{{\rm black\ dot}}=\left\{\frac{1}{12},\frac 14\right\}\,.
\eqlabel{bd}
\end{equation}
The solid blue line parameterizes the thermodynamics of cascading gauge theory 
plasma \cite{abk,ksbh}. The vertical blue dashed and dotted lines represent the nonconformality parameter $\dd$ 
of KS gauge theory at the first-order confinement/deconfinement transition,
\begin{equation}
T_{\rm deconfinement}=0.6141111(3) \Lambda\,,\qquad \dd_{\rm deconfinement}=0.2238(9)\,,
\eqlabel{ksconf}  
\end{equation}
and the (perturbative) chiral symmetry breaking phase transition,
\begin{equation}
T_{\c sB}=0.8749(0) T_{\rm deconfinement}\,,\qquad \dd_{\c sB}=0.30567(2)\,,
\eqlabel{kscsb}  
\end{equation}
correspondingly. 
Notice that while conformal invariance of cascading gauge theory plasma can be broken much more strongly 
(especially in the vicinity of the phase transitions) compare to that 
of  $\caln=2^*$ plasma, the results are of little relevance to QCD QGP --- in fact, for 
QCD temperatures $T\gtrsim 0.3$GeV both  $\caln=2^*$ and  KS plasma have very similar equations of state, 
the latter are further quite reasonable compared with lattice QCD.

\section{Relaxation in strongly coupled nonconformal plasma from holography}\label{quasi}

\begin{figure}[t]
\begin{center}
\psfrag{x}{{$\dd$}}
\psfrag{y}{{$-\Im\frac{\w}{2\pi T}$}}
\includegraphics[width=5in]{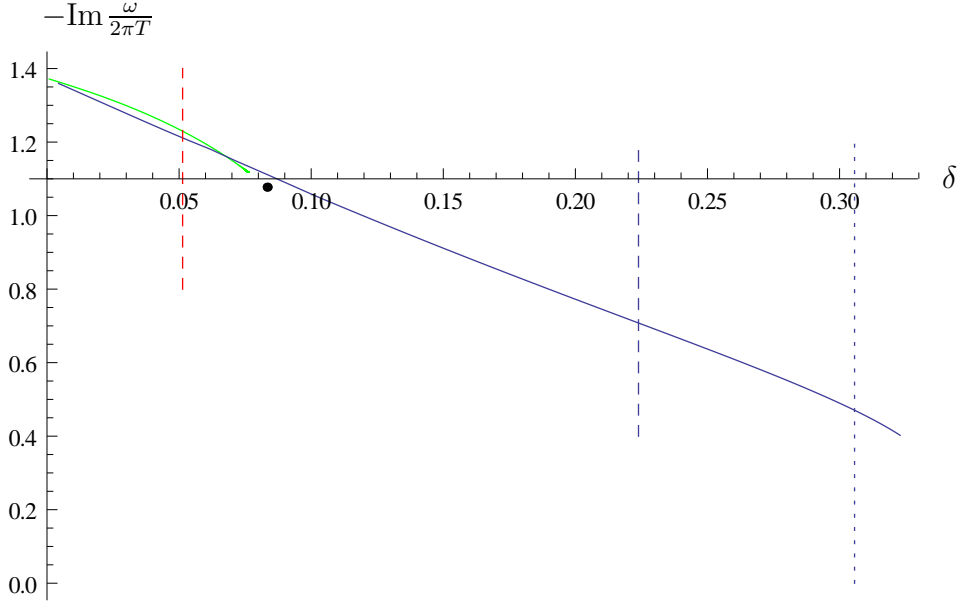}
\end{center}
  \caption{Minus imaginary part of the lowest quasinormal mode  at zero spatial momentum of the transverse traceless fluctuations 
of the stress-energy tensor in $\caln=2^*$ (the solid green line) and KS (the solid blue line) 
gauge theory plasma as a function of $\dd=\frac 13-c_s^2$.  The black dot 
denotes the lowest quasinormal mode of dimension $\Delta=5$ operator of the effective five-dimensional 
CFT in the deep IR of $\caln=2^*$ plasma, see \eqref{bdw}. The dashed red line represents the conformal violation parameter 
$\dd$ in QCD at $T=0.3$GeV.  Vertical blue lines represent the phase transitions in cascading gauge theory plasma:
the confinement/deconfinement (dashed) and the chiral symmetry breaking (dotted).} \label{figure2}
\end{figure}

\begin{figure}[t]
\begin{center}
\psfrag{x}{{$\frac{q}{2\pi T}$}}
\psfrag{y}{{$\frac{\Im\w}{\Im \w_{q=0}}$\ {\rm and}\ $\frac{\Re\w}{\Re \w_{q=0}}$ }}
\includegraphics[width=5in]{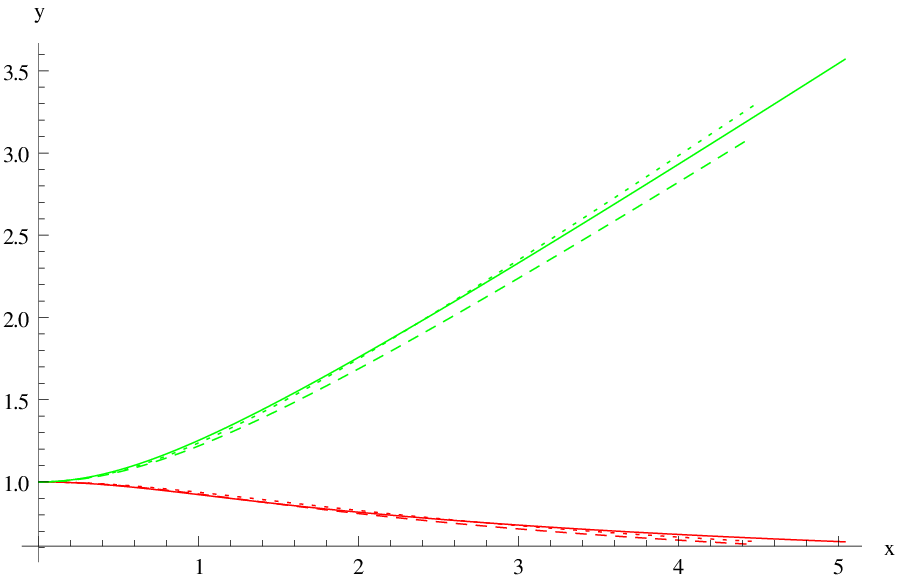}
\end{center}
  \caption{Momentum dependence of the lowest quasinormal mode of the transverse traceless fluctuations of the 
stress-energy tensor in cascading gauge theory plasma at the ultraviolet fixed point (solid lines),
the deconfinement phase transition (dashed lines), and the chiral symmetry breaking phase transition
(dotted lines). The green/red lines represent the real/minus imaginary parts of the frequencies. 
The data is normalized to zero momentum values of the frequencies, see \eqref{w0}. }\label{figure3}
\end{figure}

We now study the effects of conformal symmetry breaking on the thermalization time in
strongly coupled gauge theory plasma comparing top-down holographic models:  $\caln=2^*$ 
and KS gauge theory plasma. We focus on relaxation of the transverse traceless fluctuations 
of the stress-energy tensor. In the holographic dual they are encoded as quasinormal modes 
of helicity-2 graviton polarizations \cite{Kovtun:2005ev}. These fluctuations are always 
equivalent  to fluctuations of a minimally coupled massless scalar \cite{Buchel:2004qq}.

In figure \ref{figure2} we plot the minus imaginary part of the lowest quasinormal modes at zero spatial momentum of the transverse 
traceless fluctuations on the stress-energy tensor in $\caln=2^*$ (the solid green line) 
\cite{Buchel:2015saa} and  cascading (the solid blue line) gauge theory plasma
as a function of the conformal symmetry breaking parameter $\dd=\frac 13-c_s^2$. 
In $\caln=2^*$ gauge theory plasma $\dd\in [0,\frac{1}{12}]$ with the upper limit denoted by the black dot,
representing the imaginary part of the lowest quasinormal mode of dimension $\Delta=5$ operator in the effective 
five-dimensional CFT in the IR:
\begin{equation}
\left\{\dd,-\Im\frac{\w}{2\pi T}\right\}\bigg|_{{\rm black\ dot}}=\left\{\frac{1}{12},1.07735(7)\right\}\,.
\eqlabel{bdw}
\end{equation}
Notice that over all the parameter range of $\dd$ of $\caln=2^*$ the relaxation rates 
of $\caln=2^*$ gauge theory and KS gauge theory are practically identical. This is the basis of the 
universality observation for the relaxation rates in strongly coupled nonconformal gauge theory plasma 
with  a dual holographic description. 

In \cite{Buchel:2015saa} it was found that momentum dependence of the relaxation rates 
is rather weak in strongly coupled gauge theory plasma with a holographic dual. 
We confirm that observation here comparing the momentum dependence of the 
lowest quasinormal mode of the transverse traceless fluctuations of the stress-energy 
tensor in KS plasma for three value of $\dd$: $\dd=0$ (the UV conformal fixed point) (solid lines),
$\dd=\dd_{{\rm deconfinement}}$ (dashed lines), and $\dd=\dd_{\c sB}$ (dotted lines), see \eqref{ksconf} and \eqref{kscsb}. 
In figure \ref{figure3} we plot real (green) and minus imaginary (red) parts of the quasinormal frequencies, 
reduced to their zero momentum values:
\begin{equation}
\begin{split}
&\w_{q=0}^{\dd=0}=1.5597(3)-i\ 1.3733(4)\,,\\
&\w_{q=0}^{\dd=\dd_{{\rm deconfinement}}}=1.5825(8)-i\ 0.70783(5)\,,\\
&\w_{q=0}^{\dd=\dd_{\c sB}}=1.46632-i\ 0.47044(1)\,.
\end{split}
\eqlabel{w0}
\end{equation}

\section{Conclusion}\label{conclude}
Relaxation rates in strongly coupled gauge theory plasma are encoded in the 
lowest quasinormal modes of matter-gravity fluctuations in the corresponding holographic dual.
We studied the dependence of the relaxation rates on the mechanism of breaking the conformal 
invariance in top-down holographic models. Specifically, we compared $\caln=2^*$ gauge theory 
rates \cite{Buchel:2015saa} with those of cascading gauge theory. In the former,
the conformal invariance is broken by relevant operators, while in the latter it is broken 
by marginal (but not exactly marginal) operators. Remarkably, at least for the relaxation of 
transverse traceless fluctuations of the stress-energy tensor, the rates are very close. 
Additionally, we found very weak momentum dependence of the quasinormal mode frequencies. 
All these provide further support for the universality of the relaxation rates in 
strongly coupled gauge theories with holographic duals  observed in 
\cite{Buchel:2015saa,Fuini:2015hba,Janik:2015waa,Ishii:2015gia}. 
 
It is important to emphasize that not all relaxation rates in cascading gauge theory plasma 
are roughly proportional to the temperature. For example, in the vicinity of the chiral symmetry 
breaking phase transition, the symmetry breaking fluctuations destabilize the system \cite{ksbh}
with the corresponding relaxation  rate vanishing precisely at the transition point. 
We believe that this subtlety is of little consequence to QCD applications though, as conformal invariance 
there is broken much more strongly than in QGP  produced at RHIC and LHC (see figure \ref{figure1}).

~\\
\section*{Acknowledgments}
We would like to thank  Michal Heller and Pavel Kovtun for valuable discussions.
AB thanks the Galileo Galilei Institute for Theoretical Physics for the hospitality 
and the INFN for partial support during the completion of this work.
Research at Perimeter
Institute is supported by the Government of Canada through Industry
Canada and by the Province of Ontario through the Ministry of
Research \& Innovation. AB gratefully acknowledge further support by an
NSERC Discovery grant.

\end{document}